\documentclass[aps,twocolumn,prd,epsf,showpacs]{revtex4}

\newcommand{\be}{\begin{equation}}
\newcommand{\ee}{\end{equation}}
\def\bq{\begin{eqnarray}}
\def\eq{\end{eqnarray}}
\def\beq{\begin{eqnarray}}
\def\eeq{\end{eqnarray}}

\def\ba{\begin{eqnarray}}
\def\ea{\end{eqnarray}}

\usepackage{psfig}
\usepackage{graphicx}
\usepackage{dcolumn}
\usepackage{bm}

\newcommand{\s}{\ensuremath{\psi(t,r)}}
\newcommand{\n}{\ensuremath{\nu(t,r)}}

\newcommand{\M}{\ensuremath{{\cal M}}}

\begin{document}

\title{A resolution of spacetime singularity and black hole paradoxes
through avoidance of trapped surface formation in Einstein gravity}      

\author{Pankaj S. Joshi} 
\email{psj@tifr.res.in}
\author{Rituparno Goswami}
\email{goswami@tifr.res.in}
\address{Tata Institute of Fundamental Research,\\ Homi Bhabha Road,\\ 
Mumbai 400 005, India}

\begin{abstract} The occurrence of a spacetime singularity indicates
the breakdown of Einstein gravitation theory in these extreme regimes. 
We consider here the singularity issue and various black hole paradoxes 
at classical and quantum levels. It is pointed out that a 
possible resolution to these problems could be arrived at by avoiding the 
formation of trapped surfaces during a continual gravitational collapse. A 
class of perfect fluid collapse models is constructed which realizes such 
a possibility. While the pressure could be negative in the interior of 
the cloud, the weak energy condition is satisfied. The collapsing star 
radiates away most of its matter as the process of gravitational collapse
evolves, so as to avoid the formation of trapped surfaces and the spacetime 
singularity. The collapsing interior is matched to an exterior which 
is a generalized Vaidya spacetime to complete the model.   

\end{abstract}

\pacs{04.20.Dw, 04.70.-s, 04.70.Bw}
 
\maketitle

\section{introduction}

While the Einstein gravity has been highly successful theory of
gravitation, it is well-known that it generically admits the 
existence of spacetime singularities. These are extreme regions
in the spacetime where densities and spacetime curvatures typically 
blow up and the theory must breakdown. Such singularities may develop 
in cosmology, indicating a beginning for the universe, or in a 
gravitational collapse which ensues when a massive star exhausts its 
nuclear fuel and undergoes the process of continual collapse. It is  
expected that a future possible theory of quantum gravity may resolve 
these singularities where all known laws of physics breakdown. While 
there have been very many attempts in this
direction over past decades, we still must wait for a consistent
quantum gravity theory yet to be developed, before such a possibility 
of singularity resolution through the same can be realized.

The singularity theorems predicting the occurrence of the 
spacetime singularities,  
however, contain three main assumptions under which the existence of
a singularity is predicted in the form of geodesic incompleteness 
in the spacetime (see e.g.
\cite{HE}). 
These are in the form of a typical causality condition which
ensures a suitable and physically reasonable global structure of the   
spacetime, an energy condition which requires the positivity of energy
density at the classical level as seen by a local observer, and finally 
a condition demanding that trapped surfaces must exist in the dynamical 
evolution of the universe, or in the later stages of a 
continual gravitational collapse. A trapped region in the spacetime
is that consisting of trapped surfaces, which are 2-surfaces such that
both ingoing as well as outgoing wavefronts normal to the same 
must converge. Such trapped surfaces then necessarily give rise to a 
spacetime singularity either in gravitational collapse 
or in cosmology.

For the same reason, the process of trapped surface formation 
in gravitational collapse is also central to the black hole physics. 
The role of such a trapping within the framework of Einstein's theory of 
gravitation was highlighted by Oppenheimer, Snyder, and Dutt (OSD)
\cite{osd}
within the context of continual collapse of a massive matter cloud. 
They studied the collapse of a pressureless dust model 
using general relativity, and showed that it leads to the formation of 
an {\it event horizon} and a {\it black hole} as the collapse end state, 
assuming that the spatial density distribution within the star was 
strictly homogeneous, that is, $\rho=\rho(t)$ only. In the later stages 
of collapse an {\it apparent 
horizon} develops, which is the boundary of the trapped region,
thus giving rise to an event horizon. Once 
the collapsing star has entered the event horizon, the causal
structure of spacetime there would then imply that no non-spacelike 
curves from that region would escape away, which is cut off from the 
faraway observers, thus giving rise to a black hole in the spacetime.

While black hole physics has led to several exciting 
theoretical as well as observational developments, it is necessary, 
however, to study more realistic models of gravitational collapse 
in order to put the black
hole physics on a sound footing. This is because the OSD scenario 
is rather idealized and pressures would play an important role in the 
dynamics of any realistic collapsing star. Also density distribution
could not be completely homogeneous in any physically realistic model, 
but would be higher at the center of the cloud, with a typical negative 
gradient as one moved away from the center. The study of dynamical 
collapse within Einstein's gravity is, however, a difficult subject 
because of the non-linearity of the Einstein equations. It was hence 
proposed in 1969 by Penrose 
\cite{pen}
that any physically realistic continual gravitational collapse must  
{\it necessarily} end in a black hole final state only.
This is known as the {\it cosmic censorship conjecture}(CCC). Though CCC
has played a rather crucial role as a basic assumption in all the 
physics and astrophysical applications of black holes so far, no proof or 
any mathematically 
rigorous formulation for the same is available as yet despite many
attempts. Hence this has been 
widely recognized as the single most important problem in the theory of
black holes and the gravitation physics today. What is really necessary
to make any progress on CCC, or to understand the final endstate of
a massive collapsing star is to study realistic gravitational collapse
scenarios within the framework of Einstein's gravity
\cite{psjbook}.

Such a study of gravitational collapse is also warranted 
because of the deep paradoxes that are associated with the black holes,
and which have been widely discussed. Firstly,   
all matter entering a black hole, must, of necessity collapse into a 
spacetime singularity of infinite density and curvatures at the center
of the cloud, where all known laws of physics must break down. It is not 
clear how such a model can be stable at the classical level.
Secondly, we need to formulate mathematically and prove CCC, 
that a generic gravitational collapse gives rise to a black hole only.
However, there have been several detailed collapse studies so far which
show that the final fate of a collapsing star could be either a 
{\it black hole} or a {\it naked singularity}, depending on the nature
of initial data from which the collapse evolves, and the possible 
evolutions as allowed by the Einstein equations (see e.g.
\cite{rev} 
for some recent reviews). As opposed to a black hole final state, 
naked singularity
is a scenario where ultra-strong density and curvature regions of 
spacetime forming as result of collapse would be visible to faraway
observers in the spacetime, in violation to CCC. 
Finally, it is well-known that a black hole 
would create information loss, violating unitarity principle, thus creating
contradiction with basic principles of quantum theory.

Our purpose here is to investigate the possibility that
the singularity problem as well as black hole paradoxes 
can be resolved by possibly avoiding the trapped surface formation
in the spacetime during the process of a dynamical gravitational collapse.
We construct here a class of collapsing solutions for the Einstein
equations, where the trapped surface formation could be delayed 
or avoided during the collapse. This could then offer a somewhat natural
solution to above issues within the framework detailed below. 
In the models that we develop, much of the matter of the star can 
escape away and would be thrown out during final stages of gravitational 
collapse, thus resolving the problems such as above
and also that of the infinite density spacetime singularity. Towards
realizing such a scenario, we impose various physical reasonability 
conditions. While we require regularity of the initial data as well
as weak energy condition, we allow pressure to be negative. 
We show that in the class of perfect fluid collapse solutions we
develop, no trapped surfaces form as the collapse evolves in 
time, thus allowing the matter to escape away in the later stages of 
gravitational collapse. The perfect fluid form of matter is chosen as
it could admit several physically realistic equations of state,
and it is a form of matter which
has been studied extensively within astrophysical contexts.

The next section 2 specifies the basic regularity conditions on 
the class of models and the Einstein equations to be solved. 
In the section 3, the perfect fluid solutions are given where we 
show how the trapped surface formation is naturally avoided as the 
collapse progresses. To complete the model, we match this interior
collapsing spacetime to an exterior Vaidya solution in section 4.
Section 5 gives some concluding remarks.

\section{Einstein Equations and Regularity conditions}

For the class of solutions to the Einstein field equations 
that we construct here, at the initial epoch we take the matter 
to be a regular {\it isentropic} perfect fluid with a linear equation 
of state $p=k\rho$. Then the 
energy conditions imply that at the initial surface, the fluid has 
a positive pressure. As the collapse evolves, we no longer 
impose the condition on the perfect fluid to remain necessarily
to be isentropic, but allow the evolution of equation of state as 
determined by the collapse itself and the Einstein equations, 
as we shall discuss. The pressure decreases 
monotonically as the collapse proceeds, 
and eventually and closer to the singularity it necessarily becomes 
negative in this class of solutions. It is seen that there is 
avoidance of any trapped surfaces formation in the spacetime. 
The {\it weak energy condition} is, however, preserved through out 
the collapse. The interior collapsing sphere is matched with a 
generalized Vaidya exterior spacetime, and the energy of the collapsing 
configuration is radiated away, when the model is smoothly matched 
to a flat Minkowski spacetime.

The spherical metric in a general form can be written 
in a comoving frame as
\begin{equation}
ds^{2}=-e^{2\nu(t,r)}dt^{2} + e^{2\psi(t,r) }dr^{2} + R^{2}(t,r)d\Omega^{2},
\label{eq:metric}
\end{equation}
where $d\Omega^{2}$ is the line element on a two-sphere. 
The energy-momentum tensor for a perfect fluid is given as,
\begin{equation}
T^{t}_{t}=-\rho(t,r);\; 
T^{r}_{r}= T^{\theta}_{\theta}=T^{\phi}_{\phi}=p(t,r)  
\label{eq:setensor}
\end{equation}
where $\rho$ and $p$ are the energy density and pressure 
respectively. We assume that the matter
field satisfies the {\it weak energy condition}, that is, the 
energy density as measured by any local observer is non-negative, 
and for any timelike vector $V^{i}$, we have 
\begin{equation}
T_{ik}V^{i}V^{k}\ge0 
\end{equation}
which amounts to $\rho\ge0, \rho+p\ge0$.
The dynamic evolution of the initial data is then determined by the 
Einstein equations, which for the metric (\ref{eq:metric}) become ($8\pi 
G=c=1$),
\begin{eqnarray}
\rho=\frac{F^{\prime}}{R^{2}R^{\prime}} & &
p=-\frac{\dot{F}}{R^{2}\dot{R}}  
\label{eq:ein1}
\end{eqnarray}
\begin{equation}
-p'=\nu'\left(p+\rho\right)
\label{eq:ein2}
\end{equation}
\begin{equation}
R'\dot{G}-2\dot{R}\nu'G=0
\label{eq:ein3}
\end{equation}
\begin{equation}
G-H=1-\frac{F}{R},  \label{eq:ein4}
\end{equation}
The functions $G$ and $H$ above are defined as 
\begin{equation}
G(t,r)=e^{-2\psi}(R^{\prime})^{2}, H(t,r)=e^{-2\nu} (\dot{R})^{2}
\end{equation}
The arbitrary function $F=F(t,r)$ here has an 
interpretation of the mass function for the cloud, and it
gives the total mass in a shell of comoving radius 
$r$ on any spacelike slice $t=const$. The energy conditions imply 
$F\ge0$. 
To preserve regularity at the initial epoch, we must have  
$F(t_i,0)=0$, that is, the mass function should vanish at the center 
of the cloud.

We shall like to consider here only the collapsing
class of solutions to the Einstein equations because we are
interested in the issues related to the spacetime singularity. 
This implies that we have $\dot R<0$, that is, the physical radius $R$ of the 
cloud decreases in time. In a continual collapse scenario this
ultimately reaches $R=0$, which is the singularity where all
matter shells collapse to a zero physical radius.
We use the scaling freedom for the radial coordinate 
$r$ to write $R=r$ at the initial epoch $t=t_i$, and introduce a 
function $a(t,r)$ as defined by,
\begin{equation}
a(t,r)\equiv R/r 
\label{eq:R}
\end{equation}
We then have 
\begin{equation}
R(t,r)=ra(t,r),\; a(t_i,r)=1,\; a(t_s(r),r)=0 
\end{equation}
with $\dot{a}<0$.
The time $t=t_s(r)$ corresponds to the shell-focusing 
singularity at $R=0$, where all the matter shells collapse to a 
vanishing physical radius. Now, from equation (\ref{eq:ein2}), we
see that the equation of state is given implicitly as,
\begin{equation}
p=-\rho+e^{-\nu}\int \rho'e^{\nu}dr+\beta(t)e^{-\nu}
\label{eq:eos}
\end{equation}

We note that there are a total of five field equations 
with six unknowns, $\rho$, $p$, 
$\psi$, $\nu$, $R$, and $F$, giving us the freedom of choice of
one free function. Selection of this function, subject to the given
initial data and the weak energy condition, determines the matter 
distribution
and metric of the spacetime, and thus leads to a particular 
dynamical collapse
evolution of the initial data.
Also, the regularity conditions (e.g. density should not blow up 
at the regular center of the cloud at any epoch before the 
singularity) imply that 
at any non-singular epoch we have necessarily $F\approx r^3$  
near the center, which fixes the general form of $F$ as
\begin{equation}
F(t,r)=r^3\M(r,a)
\label{eq:mass}
\end{equation}
where $\M(r,a)$ would be any general function. It is to be
noted that this is no special choice, but the general form as 
determined by the regularity conditions.

\section{Collapsing perfect fluid models}

As pointed out above, towards constructing any class of perfect
fluid solutions to the Einstein equations, we have the freedom to 
fix one free function. We choose this to be the mass function
$F(t,r)$ for the collapsing cloud. At any given epoch, the cloud has
a compact support on a spacelike surface of $t=const.$, outside
which the collapsing ball is matched to a suitable exterior spacetime
as we shall show in the next section.

Towards this purpose, we first give below a
specific model where a definite choice of the mass function is 
being made as under,
\begin{equation} 
\M=M_0[5a^2-4a^3]
\label{eq:massexample}
\end{equation}
Clearly, this is a mass function with the required
differentiability as needed by Einstein equations and we
have $\M(r,a)\ge0$. Also, in the limit as $a\to0$, the
mass function goes to a vanishing value as a power of $a(r,t)$,
and finally there is a definite value $a=a^*$ where 
the derivative $\M,{_a}$ changes sign which corresponds to $a^*=5/6$.

With the choice of the mass function as above, let us define an
arbitrary function $A(r,a)$ as, $A(r,a)_{,a}\equiv \nu'/R'$.
Now from equation (\ref{eq:ein2}) we can easily see that, 
\begin{equation}
A(r,a)_{,a}=-\frac{2a'}{a^2}
\label{eq:Aexample}
\end{equation}
With the form of $A$, we can now integrate equation
(\ref{eq:ein3}) to get,
\begin{equation}
G(r,a)=[1+r^2b_0(r)]e^{2rA}
\end{equation}
where $b_0(r)$ is an arbitrary function of integration.
It can now be checked by direct substitution that
any $A$ and $b_0(r)$, of the form
\begin{equation}
A=rf(r);\;\; b_0(r)=\frac{(1+\alpha r^2)e^{-rA}-1}{r^2}
\label{eq:Aexample1}
\end{equation}
where $\alpha$ is a constant,
consistently solves the system of Einstein equations,
with $a=a(t)$, which is given in the integral form, 
by the equation of motion as,
\begin{equation}   
t=\int_a^1\frac{\sqrt{a}da}{\sqrt{a\alpha+M_0[5a^2-4a^3]}}
\label{eq:texample1}
\end{equation}
Now from equation (\ref{eq:ein1}), we get the density and pressure
as,
\begin{equation} 
\rho=3M_0\left[\frac{5}{a}-4\right]
\label{eq:rhoexample1}
\end{equation}
and,
\begin{equation} 
p=2M_0\left[6-\frac{5}{a}\right]
\label{eq:pexample1}
\end{equation}
It can be easily seen now that the weak energy condition is
satisfied throughout the collapse as we have $\rho>0$ and 
$\rho+p>0$. Furthermore we have an equation of state of the
form $p=k(t)\rho$ where the function $k(t)$ is given as,
\begin{equation} 
k(t)= \frac{12a-10}{15-12a}
\end{equation}
Thus we see that, at the initial hypersurface the pressure
is positive, but as the collapse proceeds the pressure decreases and
ultimately becomes negative near the singularity.

The above thus provides the full solution to the Einstein equations.
In order to see that this represents a continual collapse model,
it is important to note that at the initial epoch
we are imposing the collapse condition {\it i.e.} $\dot{R}<0$.
However, in principle, it could so happen that the collapsing system
bounces back at a later epoch before actually reaching the singularity. 
A necessary
condition for such a bounce to take place is $\dot{a}=0$ for
some value $a\in(0,1)$. But from the equation
(\ref{eq:texample1}) it is evident that if we take the constant
$\alpha$ to be positive, which is a free parameter, then
$\dot{a}<0$ holds throughout the dynamical evolution
of the model. That is, in that case collapse would be continual without
any bounce. Thus in this case, the suitable choice of this constant
specifies continual collapse models.

Generalizing the above specific model now, 
let us consider the following class of mass functions $F(t,r)$
for a collapsing cloud, where $\M(r,a)$ is taken to be an arbitrary 
function with the following properties,

\begin{enumerate}
\item $\M(r,a)$ is a $C^2$ function and $\M(r,a)\ge 0$
\item $\lim_{a\rightarrow 0}$ $\M$ goes to zero as $a^\alpha$, where
      $1<\alpha<3$
\item There exists a value $a^*$, in the interval $(0,1)$ such that, 
      $\M_{,a}|_{a>a^*}<0$, $\M_{,a}|_{a<a^*}>0$ and $\M_{,a}|_{a=a^*}=0$.
\end {enumerate} 
Then equation (\ref{eq:ein1}) gives,
\begin{equation}  
\rho=\frac{3\M+r\left[\M_{,r}+\M_{,a}a'\right]}{a^2(a+ra')};\;p=-\frac{\M_{,a}}
{a^2}
\label{eq:p}
\end{equation}

Apart from motivating our discussion, the specific
model given above shows that the solution set of Einstein equations 
for this class of mass functions with properties as above is 
non-empty. Clearly, for $a>a^*$ the pressure is positive, and it
changes sign for $a<a^*$. At $a=a^*$ the matter is dustlike with zero
pressure. The energy conditions put some further restrictions on $\M$,
namely,
\begin{equation}
3\M+r\left[\M_{,r}+\M_{,a}a'\right]\ge 0,\; 3\M+r\M_{,r}-a\M_{,a}\ge0
\end{equation}
Furthermore, if at the initial epoch the perfect fluid is isentropic
with a linear equation of state $p=k\rho$, then an initial condition
on the mass function is,
\begin{equation} 
3k\M(r,1)+kr\M(r,1)_{,r}+[\M_{,a}]_{a=1}=0
\label{eq:ec3}
\end{equation}
Now using the mass function $\M$ equation (\ref{eq:ein2})
becomes,
\begin{equation} 
\frac{\nu'}{R'}=\frac{a\M_{,ra}+a'[\M_{,aa}a-2\M_{,a}]}
{a[3\M+r\M_{,r}-a\M_{,a}]}
\label{eq:A}
\end{equation}
As earlier, let us define  
\begin{equation}
\nu'/R' \equiv A(r,a)_{,a}
\end{equation}
From the regularity
conditions, and the requirement that the gradients of pressure 
vanish at the center 
of the cloud, one can see that $A\approx rq(r,a)$ near the center, 
where $q$ is any regular function. Our main interest here is in 
studying the shell-focusing at $R=0$, which is the physical 
termination of collapse. Hence we assume that there are no shell-crosses 
where $R'=0$, and so the function $A(r,v)$ is well-defined.
Now using the above form of $\nu'$ we can integrate (\ref{eq:ein3}) 
to get,
\begin{equation}
G(r,v)=b(r)e^{2rA}
\label{eq:G}
\end{equation}
Here $b(r)$ is another arbitrary function of 
comoving coordinate $r$. A comparison with dust 
collapse models interprets $b(r)$ as the 
velocity function for the collapsing shells
\cite{gj}.
Following this parallel, 
we can write 
\begin{equation}
b(r)=1+r^2b_0(r)
\end{equation}
Thus we see that the metric (\ref{eq:metric}) becomes
\begin{equation}
ds^{2}=-e^{2\int A_{,a}R'dr}dt^{2}+ \frac{R'^2e^{-2rA}}{[1+r^2b_0(r)]}dr^{2} 
+ R^{2}(t,r)d\Omega^{2},
\label{eq:metric2}
\end{equation}
Finally, from equation (\ref{eq:ein4}) we get,
\begin{equation}
\sqrt{a}\dot{a}=-e^\nu\sqrt{e^{2rA}ab_0(r)+ah(r,a)+\M(r,a)}
\label{eq:collapse1}
\end{equation}
where the negative sign depicts a collapse scenario, and,
\begin{equation}
h(r,a)= \frac{e^{2rA}-1}{r^2}
\end{equation}
Integrating the above we get,
\begin{equation}
t(a,r)=\int_a^1\frac{\sqrt{a}da}{e^\nu\sqrt{e^{2rA}ab_0+ah+\M}}
\label{eq:scurve1}
\end{equation}
Note that the variable $r$ is treated as a constant in the above equation. 
Now putting $a=0$ in the above equation would give us the
time for termination of collapse $t=t_s(r)$, {\it i.e.} the time 
taken for the
shell labeled $r$ to reach the vanishing physical radius $R=0$,
\begin{equation}
t_s(r)=\int_0^1\frac{\sqrt{a}da}{e^\nu\sqrt{e^{2rA}ab_0+ah+\M}}
\label{eq:scurve2}
\end{equation}

In general of course, given a rather general 
choice of the mass function with the properties as given above, 
it may not be always possible to find in an explicit manner 
the complete solution to the
system of Einstein equations, with the area radius function 
$a=a(t,r)$ having a general form. 
We can, however, find an approximate solution near the center ($r=0$), 
by Taylor expanding the right hand side of the equation (\ref{eq:scurve1}) 
around $r=0$ (see for example
\cite{gj}), 
by using equation (\ref{eq:A}) and the form of $\nu$ as above. 
By doing so we can see how the collapse terminates and
behaves in its later stages depending upon the initial profiles 
and the allowed dynamic evolutions.

The key factor that would now decide the final outcome 
of the collapse in terms of a spacetime singularity and black hole 
formation or otherwise, 
is the geometry of trapped surfaces which may develop as the collapse
evolves. These are
two-surfaces in the spacetime geometry from which both outgoing and 
ingoing wavefronts necessarily converge
\cite{HE}.
If the trapped surfaces necessarily formed as the collapse
develops, then a region arises in the spacetime where any
particle or light ray entering the same will not be able to
escape away, and must necessarily fall into a spacetime singularity.
In general, the existence of trapped surfaces, together with
reasonable causality assumptions and an energy condition ensures
that a spacetime singularity must develop where the spacetime curvatures
and other physical quantities blow up, and where the known laws of 
physics must break down. This is the main content of the 
singularity theorems which apply to both gravitational collapse 
as well as cosmology situations. As we discussed 
above, the CCC would demand that any spacetime singularities 
forming in gravitational collapse must necessarily be hidden
within a black hole, that is, the extreme strong curvature regions 
should not be visible to any external observers in the spacetime.
Of course, if any of the basic assumptions of the singularity
theorems do not hold, then there is a possibility in principle 
that the singularity can be avoided. While
the causality and a reasonable global structure of the spacetime
would be necessary for any physical model, and the energy conditions
may also be required to hold, the formation or otherwise of the trapped 
surfaces during the collapse would depend on the nature of dynamical
evolution of the collapsing cloud.

The boundary of the trapped
region is the {\it apparent horizon}, which signals the black hole
formation in gravitational collapse.
The apparent horizon in a spherically symmetric spacetime 
is given by,
\begin{equation}
F=R
\label{eq:apphor}
\end{equation} 
The spacetime region where the mass function $F$ satisfies $F<R$ is 
not trapped, while $F>R$ describes a trapped region (see e.g.
\cite{jd}).
Considering again the continual collapse example that we discussed
in the beginning of this section, regularity would demand that there 
are no trapped
surfaces at the initial epoch. If $r=r_b$ denotes the boundary of the
cloud, then imposing the condition $M_0 (r_b)^2 <1$ ensures that  
there will be no trapped surfaces for any of the shells at $r\le r_b$
because $F/R< 1$ is preserved at the initial epoch. In fact, we see 
that imposing a somewhat stronger condition, namely $M_0 (r_b)^2 < 16/25$ 
is sufficient to ensure that $F/R<1$ is preserved throughout till the 
termination of collapse at $a=0$.

In the general case, in the limit of termination of collapse which is
at $a=0$, we see that for the class of perfect fluid collapse models 
we have considered here, 
\begin{equation}
\frac{F}{R}\approx r^2a^{\alpha-1}=0
\label{eq:apphor2}
\end{equation}
Thus we see that even as the collapse comes to an end as 
the physical radii $a\to0$ for the collapsing shells, there
are no trapped surfaces forming in the spacetime. The reason why
this happens is that as the collapse 
progresses the pressures turn negative in the class of models 
constructed here. As seen from the Einstein equation (3), a negative 
pressure necessarily means
$\dot F$ must be negative, which implies that the mass function
has to be decreasing in time for each corresponding shell at a constant
value of comoving radius $r$. In other words, as the collapse progresses,
the mass is being radiated away at the same time, and as a result
there is never a sufficient mass within a given radius to allow for 
the formation of the trapped surfaces. Finally, as the collapse
terminates at $a=0$, we get $F=0$ for all values of $r$ within the
cloud, whereby {\it all} the mass of the cloud has been radiated
away. A discussion on the effect of negative pressures within
the framework of gravitational collapse models was also made in
\cite{witten}.

The above thus gives a class of perfect fluid solutions 
to the Einstein equations which represents a continual gravitational
collapse. The allowed mass function and other parameters of the
models are such that trapped surface formation is avoided through out
till the termination of collapse. The matter satisfies the weak
energy condition always, however pressures turn negative as the
collapse progresses. The important point is, we have here an
explicit construction of a class of models of gravitational
collapse from regular initial data, with a reasonable form of matter
and energy condition satisfied, such that trapping is avoided.

\section{Exterior spacetime and matching conditions}

To complete the model, we now need to match this class
of perfect fluid
interiors to a suitable exterior spacetime. In the following, 
we match this collapsing ball of perfect fluid to a generalized 
Vaidya exterior 
\cite{wang} 
at the boundary hypersurface $\Sigma$
given by $r=r_b$. Then the metric just inside $\Sigma$ is,
\begin{equation}
ds^2_{-}=-e^{2\n}dt^2+e^{2\s}dr^2+R^2(t,r)d\Omega^2
\label{eq:metricgen}
\end{equation}
while the metric in the exterior of $\Sigma$ is
\begin{equation}
ds^2_{+}=-\left(1-\frac{2M(r_v,v)}{r_v}\right)dv^2-2dvdr_v+r_v^2d\Omega^2
\label{eq:metricvaidya}
\end{equation} 
where $v$ is the retarded (exploding) null co-ordinate and $r_v$
is the Vaidya radius. Matching the area radius at the boundary we get,
\begin{equation}
R(r_b,t)=r_v(v)
\label{eq:radius}
\end{equation}  
Then on the hypersurface $\Sigma$, the interior and exterior metrics are
given by,
\begin{equation}
ds^2_{\Sigma-}=-e^{2\nu(t,r_b)}dt^2+R^2(t,r_b)^2d\Omega^2
\label{eq:metric3}
\end{equation} 
and
\begin{equation}
ds^2_{\Sigma+}=-\left(1-\frac{2M(r_v,v)}{r_v}+2\frac{dr_v}{dv}\right)dv^2
+r_v^2d\Omega^2
\label{eq:metric4}
\end{equation}
Matching the first fundamental form gives,
\begin{eqnarray}
\left(\frac{dv}{dt}\right)_\Sigma=\frac{e^{\nu(t,r_b)}}
{\sqrt{1-\frac{2M(r_v,v)}{r_v}
+2\frac{dr_v}{dv}}};&\left(r_v\right)_\Sigma=R(t,r_b)
\label{eq:match1}
\end{eqnarray} 

Next, to match the second fundamental form (extrinsic curvature)
for interior and exterior metrics, we note that the normal to the
hypersurface $\Sigma$, as calculated from the interior metric, is given as,
\begin{equation}
n^i_{-}=\left[0,e^{-\psi(r_b,t)},0,0\right]
\label{eq:n1}
\end{equation} 
and the non-vanishing components of the normal as derived from the 
generalized Vaidya spacetime are,
\begin{equation}
n^v_{+}=-\frac{1}{\sqrt{1-\frac{2M(r_v,v)}{r_v}+2\frac{dr_v}{dv}}}
\label{eq:n2}
\end{equation} 
\begin{equation}
n^{r_v}_{+}=\frac{1-\frac{2M(r_v,v)}{r_v}+\frac{dr_v}{dv}}
{\sqrt{1-\frac{2M(r_v,v)}{r_v}+2\frac{dr_v}{dv}}}
\label{eq:n3}
\end{equation} 
Here the extrinsic curvature is defined as,
\begin{equation}
K_{ab}=\frac{1}{2}{\cal L}_{\bf n}g_{ab} 
\label{eq:k1}
\end{equation} 
That is, the second fundamental form is the Lie derivative
of the metric with respect to the normal vector ${\bf n}$.
The above equation is equivalent to,
\begin{equation}
K_{ab}=\frac{1}{2}\left[g_{ab,c}n^c+g_{cb}n^c_{,a}
+g_{ac}n^c_{,b}\right]
\label{eq:k2}
\end{equation}\\ 
Now setting $\left[K_{\theta\theta}^{-}-K_{\theta\theta}^{+}\right]_{\Sigma}=0$
on the hypersurface $\Sigma$ we get,
\begin{equation}
RR'e^{-\psi}=r_v\frac{1-\frac{2M(r_v,v)}{r_v}+\frac{dr_v}{dv}}
{\sqrt{1-\frac{2M(r_v,v)}{r_v}+2\frac{dr_v}{dv}}}
\label{eq:match2}
\end{equation} 
Simplifying the above equation using equation (\ref{eq:match1}) 
and the Einstein equations, we get,
\begin{equation}
F(t,r_b)=2M(r_v,v)
\label{eq:match3}
\end{equation} 
Using the above equation and (\ref{eq:match1}) we now get,
\begin{equation}
\left(\frac{dv}{dt}\right)_\Sigma=\frac{e^\nu(R'e^{-\psi}+\dot{R}e^{-\nu})}
{1-\frac{F(t,r_b)}
{R(t,r_b)}}
\label{eq:match4}
\end{equation}
Finally, setting $\left[K_{\tau\tau}^{-}-K_{\tau\tau}^{+}\right]_{\Sigma}=0$,
where $\tau$ is the proper time on $\Sigma$,  
we get,
\begin{equation}
M(r_v,v)_{,r_v}=\frac{F}{2R}+\frac{Re^{-\nu}}{\sqrt{G}}\sqrt{H}_{,t}
+Re^{2\nu}\nu'e^{-\psi}
\label{eq:match5}
\end{equation}
\begin{figure}[tb]
\hspace*{0.3cm}
\psfig {file=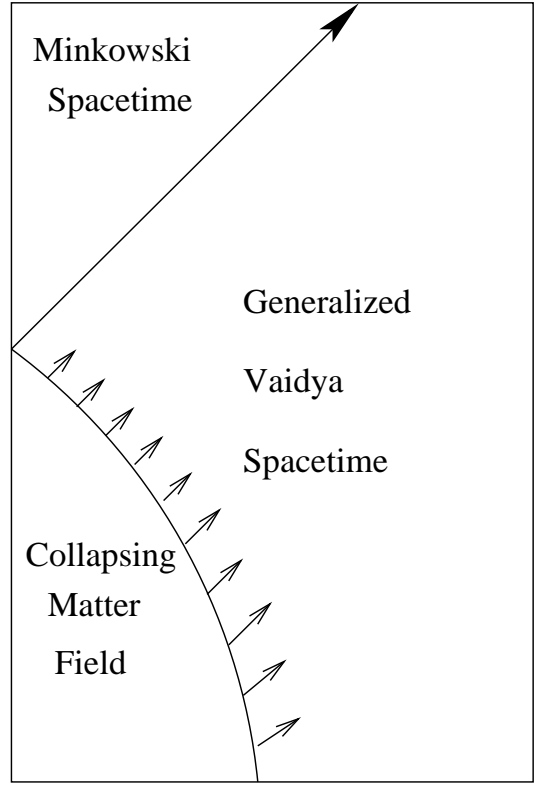,width=7cm}
\caption{A schematic diagram of the complete spacetime.}
\end{figure}

Any generalized Vaidya mass function $M(v,r_v)$, which satisfies
equation (\ref{eq:match5}) will then give a unique exterior spacetime
with required equations of motion given by other matching conditions,
(\ref{eq:match3}), (\ref{eq:match4}) and (\ref{eq:radius}). 
To give some examples of such functions $M(v,r_v)$,
we can have charged Vaidya spacetime as an exterior in which 
$M=M(v)+Q(v)/r_v$, 
or one can have an anisotropic 
de Sitter exterior where $M=M(r_v)$, which are two different solutions 
of the
equation (\ref{eq:match5}) (see for example ~\cite{wang} and 
~\cite{match}). 
This gives examples of two unique exterior spacetimes, both of which are 
subclasses
of generalized Vaidya model considered here.

Now we can easily see that along the singularity curve, $t=t_s$,
\begin{equation}
\lim_{ r_v\rightarrow 0} \frac{2M(r_v,v)}{r_v}\rightarrow 0
\label{eq:match6}
\end{equation}
Thus the exterior metric along with the singularity smoothly transform to,
\begin{equation}
ds=-dv^2-2dvdr_v+r_v^2d\Omega^2
\label{eq:metric5}
\end{equation} 
The above metric describes a Minkowski spacetime in retarded null 
coordinate.
Hence we see that the exterior generalized Vaidya metric, 
together with the singularity, 
can be smoothly extended to the Minkowski spacetime as the collapse 
completes.

\section{Conclusion}

We make several concluding remarks in this final section.

1. We have given here a class of perfect fluid collapse models, 
where beginning from an isentropic equation of state,
and initially positive pressures, the collapse evolution is such
that the pressures become negative in the later stages of evolution.
This causes the mass of the cloud to be radiated away, resulting in
non-formation of trapped surfaces all the way till the collapse
ends. The generalized Vaidya spacetime represents the exterior geometry 
of such a collapsing matter cloud. In such a case, as we have seen,
there are no trapped surfaces as there is never enough mass in a given 
spacetime radius to generate trapping of light.

2. The physical picture then would be that of a cloud radiating 
away most of its mass in the later stages of collapse. Eventually, as 
end state of collapse approaches, the entire star evaporates and the 
geometry is smoothly matched to a Minkowski spacetime.

3. In such a case, the spacetime singularity and the black hole
paradoxes are naturally resolved. Such a resolution of black hole
paradoxes might appear to be somewhat radical, in the sense
that it does away with the black hole itself. This is because as the 
trapped surface formation is avoided here, there is no event horizon
developing in the interior of the cloud and a black hole does not develop 
as the gravitational collapse end state. The point, however, is that
the Einstein equations readily admit such a possibility, as seen by the
class of solutions given here.

4. Such a scenario may deserve a serious consideration
because of the serious problems and paradoxes present at the very 
heart of the black hole physics. What we have constructed here is a 
class of perfect fluid solutions to general relativity, showing that in 
the Einstein's theory such a possibility is very much present 
emerging as gravitational collapse end state, within the framework of 
reasonable physical conditions as outlined above. It would be relevant to 
note here that in the context of cosmology, the avoidance of trapped 
surface formation has been used earlier, for a class of cylindrically 
symmetric perfect fluid cosmological models, to resolve the issue 
of spacetime singularity
\cite{seno}.

5. One could ask what would be the physical mechanism which 
could possibly give rise to a negative pressure as considered here. 
As such, in cosmology today, the existence of negative pressures is 
regarded as a fairly common phenomena, especially if a positive cosmological 
constant supplied the dark energy of the universe. In fact, such a form of dark 
energy may violate some of the energy conditions as well, as opposed to 
the weak energy condition holding in our case. However, 
within the framework of gravitational collapse, while dealing with finite 
massive matter clouds such as collapsing stars, an energy condition 
may be supposed to hold, at least at the classical level till a 
fairly advanced stage of collapse is reached. On the other hand, there 
is no problem having negative pressures, even when the weak energy 
condition is satisfied, as we have seen above. As the collapse 
progresses the quantum corrections may become important, and such
quantum effects may possibly give rise effectively to negative 
pressures, at least in a certain approximation. These issues will be 
discussed elsewhere.

\end{document}